\documentclass[a4paper]{article}
\usepackage{url}
\usepackage{graphicx}
\usepackage{rotating}
\usepackage{multirow}
\usepackage{hyperref}
\usepackage[margin=2.0cm]{geometry}

\usepackage[utf8]{inputenc}

\def\etal{{\it et al.}}

\begin{document}

\title{Insights from Machine-Learned Diet Success Prediction\thanks{This is a preprint of an article appearing at the Pacific Symposium on Biocomputing (PSB) 2016.}}
\author{Ingmar Weber\\
Qatar Computing Research Institute\\
Doha, Qatar\\
iweber@qf.org.qa
\and Palakorn Achananuparp\\
Singapore Management University\\
Singapore\\
palakorna@smu.edu.sg}

\maketitle

% ABSTRACT
\begin{abstract}
To support people trying to lose weight and stay healthy, more and more fitness apps have sprung up including the ability to track both calories intake and expenditure. Users of such apps are part of a wider ``quantified self'' movement and many opt-in to publicly share their logged data. In this paper, we use public food diaries of more than 4,000 long-term active MyFitnessPal users to study the characteristics of a (un-)successful diet. Concretely, we train a machine learning model to predict repeatedly being over or under self-set daily calories goals and then look at which features contribute to the model's prediction. Our findings include both expected results, such as the token ``mcdonalds'' or the category ``dessert'' being indicative for being over the calories goal, but also less obvious ones such as the difference between pork and poultry concerning dieting success, or the use of the ``quick added calories'' functionality being indicative of over-shooting calorie-wise. This study also hints at the feasibility of using such data for more in-depth data mining, e.g., looking at the interaction between consumed foods such as mixing protein- and carbohydrate-rich foods. To the best of our knowledge, this is the first systematic study of public food diaries.
\end{abstract}

%\keywords{MyFitnessPal; Calorie Counting; Weight Loss; Quantified Self}

%\bodymatter{}

% INTRODUCTION
\section{Introduction}\label{sec:introduction}
In 2012, 30-50 million Americans were on a diet at any given point in time\footnote{\url{https://www.npd.com/wps/portal/npd/us/news/press-releases/the-npd-group-reports-dieting-is-at-an-all-time-low-dieting-season-has-begun-but-its-not-what-it-used-to-be/}} for reasons ranging from lowering the risk of diseases to having a more positive self image. The annual revenue of the U.S. weight-loss industry is estimated at around \$20 billion.\footnote{\url{http://abcnews.go.com/Health/100-million-dieters-20-billion-weight-loss-industry/story?id=16297197}}
Clearly, dieting is not easy and new fashion diets come into existence every year.

\begin{figure}
\centering
\includegraphics[width=\columnwidth]{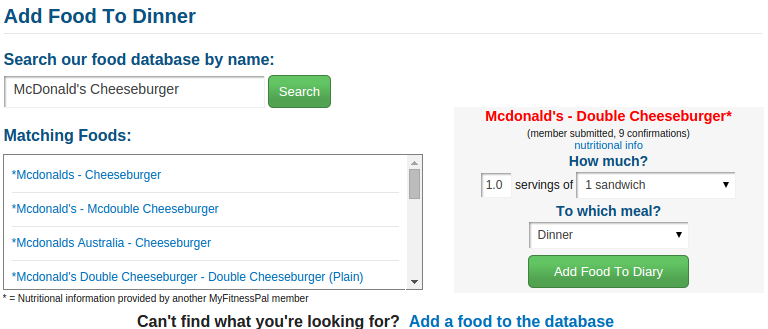}
\caption{Screenshot showing the food selection process in the MyFitnessPal web interface.}
\label{fig:food_selection}
\end{figure}

\begin{figure}
\centering
\includegraphics[width=\columnwidth]{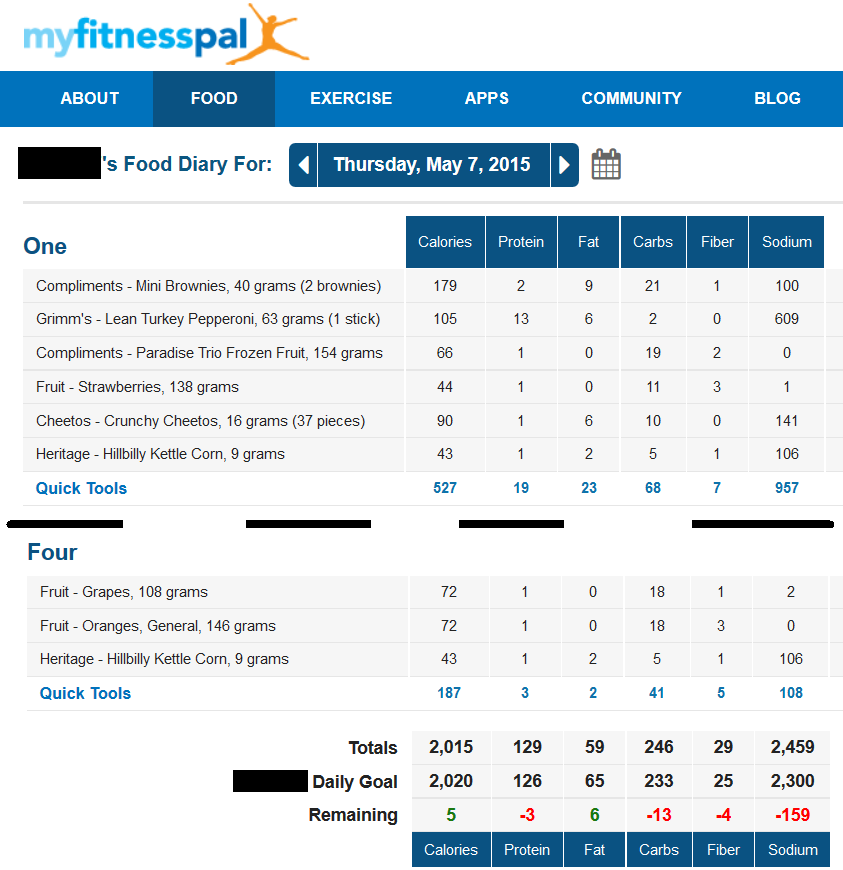}
\caption{Screenshot of a user's public food diary on MyFitnessPal. The user kept the default names for the meals (``One'', ``Two'', ...) and only the first and last meal are included in the screenshot. The bottom shows the user's actual calories consumption of 2,015kcal and their daily goal of 2,020kcal.}
\label{fig:mfp_screenshot}
\end{figure}

In this paper we explore the practice of keeping an online food diary and its relation to dieting outcomes. Concretely, we turn to data from a large fitness and health application, MyFitnessPal (henceforth MFP), and look at the publicly logged food consumption of more than 4,000 users over several months. Figure~\ref{fig:food_selection} shows the interface through which users enter their consumed food. Users not only log their daily intake but they also specify a ``daily calories goal'' against which their consumption can be compared. Figure~\ref{fig:mfp_screenshot} shows this goal at the bottom of the screenshot. Though we cannot observe their actual weight progression, we are using the information on whether a user mostly consumes more or less calories than their self-declared goal as an indicator for dieting success.

Using these user labels of being ``above'' or ``below'' we train a classifier to tell the two user groups apart using the types of food users have logged. Through a feature analysis of the classifier we gain insights into which foods are associated with diet success or failure.

Our findings are largely intuitive, e.g., logging food with ``oil'', ``butter'' or ``mcdonalds'' in the name is an indication of going above one's calorie goals. However, we also discover less obvious trends such as a distinction between pork (indicative for being ``over'') and poultry (indicative for being ``under''). In addition, we describe general behavior related to food logging. E.g., users are least likely to log \emph{any} meal on the weekend and, if they do, they are most likely to be above their weight goal.

To the best of our knowledge, this is the first systematic analysis of public food diaries. We believe that this paper helps to show the potential that this data holds for various health-related analyses.

% RELATED WORK
\section{Related Work}\label{sec:related_work}

Our research is centered around ``quantified self'' data. The quantified self is a movement to voluntarily log personally relevant data for self knowledge and improvement. Users might decide to quantify or track their activity both for goal-driven or documentary motivation \cite{rooksbyetal14chi}. Striking the right balance between the amount of data that an application wants to collect and the amount of effort required on the user's end can be challenging \cite{meyeretal14chi}. Rusin \etal \cite{rusinetal13ijmi} offer a review of technologies used for logging food intake.

When it comes to weight loss and weight maintenance, certain practices related to the quantified self such as ``think about how much progress you've made'' (by having charts showing this), ``weigh yourself'' (through weight logging) and ``read nutrition labels'' (making use of the nutrition database that services such as MFP provide) have been among the few indicators of both successful weight loss and maintenance \cite{sciamanna11ajpm}.  Other research has also found that ``self monitoring'' is important both for weight loss 
%``significant association between self-monitoring and weight loss''
\cite{burkeetal11jada} 
and for successfully maintaining such loss 
%``self-monitoring appears to be an important ongoing strategy for many of the successful losers''
\cite{kraschnewskietal10ijo}. These studies are, however, from the ``pre-smartphone era'' where calorie counting was done manually using pencil and paper. Still, they give credence to the potential benefits for both weight loss and weight management that apps such as MFP could offer. A currently ongoing study will also shed light on the effect of frequent weight control \cite{lindeetal15cct} on dieting outcomes. To date, few studies have, however, looked at the effectiveness of mobile apps for motivating health behavior changes \cite{payne15jmirmu}.

Concerning the analysis of food consumption through user-generated data, some studies have taken a public health approach and looked at regional or temporal patterns in food consumption. West \etal\ \cite{westetal13www} used web search and click-through data to study seasonal changes in the recipes people search for. Server logs from a recipe site were used in \cite{wagneretal14www,wagneretal14epj} to study regional differences in food preferences. Their angle is less health-centric and more culture- or preference-centric with a focus on whether ingredients or regional influences dominate the recipe choice. Using \emph{only} the ingredients of recipes, rather than regional or cultural knowledge, has been explored in \cite{tengetal12websci} for the purpose of recipe recommendation.

Conceptually close to our study is the work by Abbar \etal \cite{abbaretal15chi} who look at food mentions on Twitter. Their analysis also includes bit on \emph{individual} and not only public health. If users were to mention \emph{everything} they ate on social media then the type of food diaries we are using would become redundant. However, we believe that this is unlikely to be the case for many users and our data is far cleaner and more structured and comes with user-defined daily calorie goals. Culotta \cite{culotta14chi} also used Twitter data to study obesity and other health issues but from a purely aggregate, public health perspective. Kuener \etal \cite{kuebleretal13pone} use data from Yahoo Answers to look at issues of both mental and physical health of obese people. They find that ``obese people residing in counties with higher levels of BMI may have better physical and mental health than obese people living in counties with lower levels of BMI''. They do not, however, look at indicators related to the success or failure of weight loss.

As far as the idea of trying to predict dieting success is concerned, the work in \cite{lietal14group} is related. The authors study data from an online weight loss community and offer insights into different phases of usage which, potentially, could help to predict who will engage in the long-term and hence have a higher chance of achieving and maintaining weight loss. Finally, Park \etal \cite{parketal16cscw} study long-term sharing of MFP activity on Twitter. This also indirectly relates to dieting success as long-term engagement with a fitness application could be seen as beneficial for maintaining weight loss.

% DATA ACQUISITION AND PREPROCESSING
\section{Data Acquisition and Preprocessing}\label{sec:data}

% \subsection{Obtaining MFP Diaries}\label{sec:data:diaries}
\subsection{Obtaining Food Diaries and Constructing Food Taxonomy}\label{sec:data:diaries_taxonomy}
% The construction of our dataset begins with the download of public food diary pages of MFP users. To this end, we first extracted an initial list of members of the 10 most popular MFP groups. These groups represent communities of MFP users who share common health and lifestyle interests, for example, those who use a Fitbit activity tracker, those who aim to achieve a specific weight loss goal, or those who are interested in a couch-to-5K running program. In total, approximately 100K usernames were identified. Next, for each user, we retrieved up to the last 180 days of food diary pages (until March 2015 when the data collection took place). Ultimately, the food diary pages of 9,896 users are publicly accessible, resulting in 587,187 food diary pages retrieved. On average, each user has logged 59.3 days of diaries (S.D.\ = 54.6, median = 42) or 652.9 food entries in total (S.D.\ = 774, median = 366).

The construction of our dataset begins with the download of public food diary pages of MFP users. First, we extracted an initial list of 100K usernames from the 10 most popular MFP groups. Next, for each user, we retrieved up to the last 180 days of food diary pages (until March 2015 when the data collection took place). Ultimately, the food diary pages of 9,896 users are publicly accessible, resulting in 587,187 food diary pages retrieved. On average, each user has logged 59.3 days of diaries (S.D.\ = 54.6, median = 42) or 652.9 food entries in total (S.D.\ = 774, median = 366). According to the random samples (N = 200) of user profiles, the average age of users in the dataset is 36.6 years old (S.D. = 10.71). The vast majority of users are female (75\%) and reside in the United States (67\%). A small fraction of users (5.7\%) do not provide any demographic information.

When adding an entry in a food diary, users can either search for existing foods in MFP database or enter a new food description and associated nutritional values. The lack of controlled vocabulary in the food data creation causes several data integrity issues, e.g., the same foods may be described slightly different by different users. To mitigate the problems, we needed ways to group related foods into semantic categories. Thus, we built a food taxonomy by compiling lists of food-related categories and page names from Wikipedia\footnote{http://en.wikipedia.org/wiki/Category:Foods}. The taxonomy is manually organized into 18 main categories (e.g., staple food, meats, vegetables, etc.), 149 subcategories (e.g., wheat, rice, beef, etc.) and 4,233 entities, describing individual ingredients and meal types of food entries. For example, the entry ``McDonald's - Premium Sweet Chili Chicken Wrap (Grilled)'' will be annotated with the following set of \{main category: subcategory: entity\}: \{Staple foods: Wheat: Wrap\}, \{Meats: Poultry: Chicken\}, \{Preparation Methods: Grill\}, \{Fast foods: McDonald's\}.

% \subsection{Constructing a Food Taxonomy}\label{sec:data:taxonomy}
% When adding an entry in a food diary, users can either search for existing foods in MFP database or enter a new food description and associated nutritional values. The lack of controlled vocabulary in the food data creation causes several data integrity issues, e.g., the same foods may be described slightly different by different users.

% To mitigate the problems, we needed ways to group related foods into semantic categories. To achieve this, we built a food taxonomy by compiling lists of food-related categories and page names from Wikipedia\footnote{http://en.wikipedia.org/wiki/Category:Foods}. The taxonomy is manually organized into 18 main categories (e.g., staple food, meats, vegetables, fruits, or preparation methods), 149 subcategories (e.g., wheat, rice, beef, chicken, or salad) and 4,233 entities (child nodes). Specifically, our goal is to use the taxonomy as a knowledge source to help automatically annotate each entry in a food diary with categories describing its ingredients and meal types. For example, the entry ``McDonald's - Premium Sweet Chili Chicken Wrap (Grilled)'' will be annotated with the following set of \{main category: subcategory: entity\}: \{Staple foods: Wheat: Wrap\}, \{Meats: Poultry: Chicken\}, \{Preparation Methods: Grill\}, \{Fast foods: McDonald's\}.

\subsection{Data Preprocessing and Pruning}\label{sec:data:preprocessing}
% To extract categories from a food entry, we first lemmatized all entity words in the taxonomy and the food entry's text. Then, we iterated through the main categories to find the maximal exact substring match between the taxonomic entities and the food entry's text. E.g., the entity ``bean sprout'' is a match in ``Iga - bean sprouts'' but not a match in ``Sprouts - tiramisu espresso beans''. After a match had been found, the corresponding main category, subcategory, and entity were added to the annotation. The taxonomy has a reasonable coverage with respect to our dataset. Out of 632,652 unique food entries, 88\% were successfully annotated while 12\% produced an empty annotation. The causes of failed annotation include users' input errors, such as misspellings of food names (e.g., brocolli, avacado, or spinich), entering food names that are too short (e.g., ``Nestle - Fitness'') or non-English (e.g., huevos rancheros).

To extract categories from a food entry, we first lemmatized all entity words in the taxonomy and the food entry's text. Then, we iterated through the main categories to find the maximal exact substring match between the taxonomic entities and the food entry's text. E.g., the entity ``bean sprout'' is a match in ``Iga - bean sprouts'' but not a match in ``Sprouts - tiramisu espresso beans''. After a match had been found, the corresponding main category, subcategory, and entity were added to the annotation. The taxonomy has a reasonable coverage with respect to our dataset. Out of 632,652 unique food entries, 88\% were successfully annotated while 12\% produced an empty annotation. The causes of failed annotation include misspelled (e.g., brocolli) and non-English names (e.g., huevos rancheros).

To describe a user's food intake, we used both the categories described above and single-word tokens extracted from the text of the diary entry. The tokenization steps are as follows: (i) splitting on non-word characters\footnote{The usual regular expression definition of non-word characters was used, i.e.\ [\^{}a-zA-Z0-9\textunderscore{}].}, (ii) lower casing everything, (iii) only considering tokens of length at least three, and (iv) requiring all characters to be alphabetical (a-z). Furthermore, for both tokens and categories, we required that more than 500 distinct users had to use it, leaving us with 1,720 distinct tokens and 392 distinct categories. Finally, only users who had at least 30 logged days with at least one non-zero feature were considered, ignoring days with less than 100 calories logged. This left us with 5,797 users.

\subsection{Labeling Calorie Goals ``Success''}\label{sec:data:labeling}
The core of our analysis is centered around looking for differences between successful and unsuccessful users, where success is in relation to their self-declared daily calories goal. For each day, a user was assigned a label of ``below'', ``on-target'' or ``above'' depending on their calories goal and actual calories consumed as follows.
\begin{itemize}
\item below: (goal - actual) / goal $> .2$
\item above: actual $>$ goal
\item on-target: otherwise
\end{itemize}
We chose to label a (user, day) pair where the user exceeds their goal (i.e., too much consumption) by even a single calorie as ``above'' to encode the inherent asymmetry in dieting to lose weight. 
Table~\ref{tab:mfp_weekly} shows the trend of the (user, day) pairs when grouping by the day of the week and aggregating across users. As found in previous work looking at food consumption \cite{abbaretal15chi}, the weekend seems to be the worst period for dieting with (i) the highest fraction of ``above'' incidents, and (ii) the lowest number of logged days. Interestingly, web searches for recipes seem to follow the \emph{opposite} trend \cite{westetal13www}.

\begin{table}[ht]
\centering
\caption{This table shows the weekly logging trends for 9,896 users. The fraction of ``above'' increases slightly from its lowest on Mondays (19.0\%) to its highest on Saturdays (24.9\%). The number of total user-days logged also shows a drop on the weekend.\label{tab:mfp_weekly}}{
\begin{tabular}{c|ccccccc}
          & Mon & Tue & Wed & Thu & Fri & Sat & Sun \\ \hline
\% Above & 19.1\% & 20.0\% & 20.6\% & 21.0\% & 23.3\% & 24.9\% & 23.7\% \\
\% Ontarget & 33.5\% & 34.1\% & 33.6\% & 32.9\% & 29.9\% & 29.2\% & 31.3\% \\
\% Below & 47.4\% & 46.0\% & 45.7\% & 46.0\% & 46.8\% & 46.0\% & 44.9\% \\ \hline
\# Total & 93.0k & 91.7k & 88.7k & 85.4k & 80.7k & 73.4k & 73.6k  
\end{tabular}}
\end{table}

%\begin{figure}[ht]
%\centering
%\includegraphics[width=\columnwidth]{MFP_weekly_logging_morale}
%\caption{This graph shows the weekly logging trends for 9,896 users. The fraction of ``above'' increases slightly from its lowest on Mondays (19.0\%) to its highest on Saturdays (24.9\%). The number of total user-days logged also shows a drop on the weekend.}
%\label{fig:mfp_weekly}
%\end{figure}

To have a single label for each user, the (user, day) label pairs were aggregated across days by taking the modal class, i.e., the biggest class. Note that this means a user could have as little as 34\% of their days belonging to this class. Having a single label for each user, rather than modeling each (user, day) pair separately, had the advantages of (i) reducing noise due to the larger data set being considered, and (ii) interlinking behavior across days so that even ``good'' behavior on one day could be predictive of ``bad'' behavior on another day.

Using single-word tokens as features, with the above definition of user labels, 3,320 users were labeled ``below'', 1,546 were ``on-target'' and 931 ``above''. To have a clearer distinction between the classes, we chose not to consider users in the ``on-target'' group for further analysis. This left us with 4,251 users for the token-based analysis and ${}^{*}$ for the category-based analysis.

\begin{table}[ht]
\centering
\caption{Basic characteristics of our below-vs.-above data set containing 4,251 users after pruning (see text).\label{tab:dataset}}{
\begin{tabular}{|l|ccccc|}
\hline
                           & min & 10\% & median & 90\% & max \\ \hline
Total days logged per user        & 30  &  36  &  77    & 168  & 186 \\
\% days ``above'' per user & 0   &  1\% & 14\%   & 53\% & 95\% \\
\% days ``below'' per user & 0   & 18\% & 56\%   & 86\% & 100\% \\\hline
\end{tabular}}
\end{table}

\section{Results}\label{sec:results}

\subsection{Exploratory Cluster Analysis}

To gain a better understanding of the food consumption patterns, we performed a cluster analysis. Each of the 4,251 users for the token-based representation was mapped to a feature vector where each of the 1,720 dimension counted on how many distinct days a user used a specific token. These vectors were then normalized in 2-norm. We used X-Means \cite{pellegmoore00icml} as implemented in Weka \cite{halletal09sigkdd} as the clustering algorithm.\footnote{The exact parameters used were: Scheme:weka.clusterers.XMeans -I 2 -M 1000 -J 1000 -L 2 -H 10 -B 1.0 -C 0.5 -D ``weka.core.EuclideanDistance -R first-las'' -S 10 } The algorithm automatically chose $k=6$ as the optimal number of clusters (searching between $k=2$ and $k=10$).

\begin{table}[ht]
\centering
\caption{Summary of an XMeans clustering for the token-based normalized feature representation.\label{tab:clustering}}{\tiny
\begin{tabular}{lcccccc}
            &     Cluster 1 (n=545)  &  Cluster 2 (n=411)   &  Cluster 3 (n=686)     &  Cluster 4 (n=601)   &    Cluster 5 (n=878) & Cluster 6 (n=1,130)  \\
            & below/above & below/above & below/above & below/above & below/above & below/above \\
            & 375 (69\%) / 170 (31\%) & 304 (74\%) / 107 (26\%) & 575 (84\%) / 111 (16\%) & 460 (76\%) / 141 (24\%) & 716 (82\%) / 162 (18\%)  & 890 (79\%) / 240 (21\%) \\ \hline
\multirow{10}{*}{\begin{sideways}Biggest Rank Gains\end{sideways}}  & skimmed (+917) & grounds (+550) & creamer (+348) & coffee (+43) & great (+121) & almond (+84)\\
            & sainsbury (+856) & brewed (+412) & packet (+83) & sandwich (+34) & value (+99) & organic (+75)\\
            & semi (+832) & creamer (+361) & coffee (+74) & sausage (+34) & kraft (+59) & protein (+64)\\
            & asda (+794) & from (+118) & protein (+58) & wheat (+28) & wheat (+40) & coffee (+49)\\
            &  tesco (+621) & packet (+76) & tsp (+50) & pizza (+28) & turkey (+30) & yogurt (+32)\\
            & tea (+139) & coffee (+74) & sugar (+38) & turkey (+23) & cheddar (+26) & vanilla (+25)\\
            & coffee (+46) & tsp (+52) & vanilla (+36) & chips (+23) & light (+25) & natural (+22)\\
            & light (+20) & sugar (+32) & free (+32) & bacon (+17) & yogurt (+25) & banana (+15)\\
            & banana (+17) & free (+30) & yogurt (+31) & peanut (+14) & free (+22) & eggs (+14)\\
            & free (+16) & vanilla (+24) & natural (+22) & homemade (+12) & peanut (+17) & peanut (+13)\\\hline
\end{tabular}}
\end{table}

Table~\ref{tab:clustering} shows a summary of the obtained clusters. The cluster sizes are reasonably balanced, ranging from 411 to 1,130.  To understand the distinctive features for each cluster we looked at the tokens that went up the most in the ranking, compared to the global average. For example, a token that is globally ranked 1,000th in terms of its average user weight but is ranked 20th within a particular cluster has moved up 980 ranks. Furthermore, we required that the token had to end up in the top 40. This was done to ensure that the token is relatively frequent in the end. Note that the bigger clusters (Clusters 5 \& 6) are closer to the ``average'' and so the relative change in ranking is smaller for them.

The fraction of ``below'' and ``above'' users shows moderate variations, ranging from 69\% ``below'' to 84\%. However, the discriminative tokens do not necessarily tell an intuitive story as, e.g., Cluster 4 with ``pizza'' and ``bacon'' has a \emph{lower} fraction of ``above'' than Cluster 6 with ``organic'' and ``natural''. Clusters 2 and 3 also seem similar in that they both have many discriminative tokens related to coffee. However, Cluster 2 has an ``above'' percentage that is 10\% above that of Cluster 3.

Overall, the unsupervised clustering did not yield practical insights as clusters seemed to be more influenced by things such as brand names or shopping at particular supermarket chains, than by healthy-vs.-unhealthy food categorization. Hence, we decided to look at \emph{supervised} methods instead, where the above-vs.-below labels are central.

\subsection{Above-vs.-Below Machine Classification}

To understand the differences in food consumption between our above and below users, we trained a Support Vector Machine (SVM) classifier with a linear kernel and the default settings in the SVM-light\footnote{\url{http://www.cs.cornell.edu/people/tj/svm_light/}} implementation \cite{svmlight}. We used the same general setup for both feature sets, tokens and categories, though the dimensionality of the feature space differed (1,720 vs.\ 392). In both cases, we trained the classifier on a \emph{balanced} training set with a equal number of above and below instances, 931 for tokens and 919 for categories. The training set was then split into 10 folds, each with a 90\%-10\% train-test split. Table~\ref{tab:performance} summarizes the performance for the binary classifier.

\begin{table}[ht]
\centering
\caption{ Linear SVM classification results for the above-vs.-below calories target prediction in a 10-fold cross validation setting. The $\pm$ indicate standard deviations across 10 folds.\label{tab:performance}}{
\begin{tabular}{|lccc|}
\hline
Features & Accuracy & Precision & Recall \\ \hline
tokens & 67.3\% $\pm2.9\%$ & 67.8\% $\pm3.3\%$ & 66.1\% $\pm5.1\%$ \\
categories & 64.7\% $\pm3.9\%$ & 64.8\% $\pm4.1$ & 64.8\% $\pm4.3$ \\ \hline
\end{tabular}}
\end{table}

\begin{figure}[ht]
\centering
\includegraphics[width=\columnwidth]{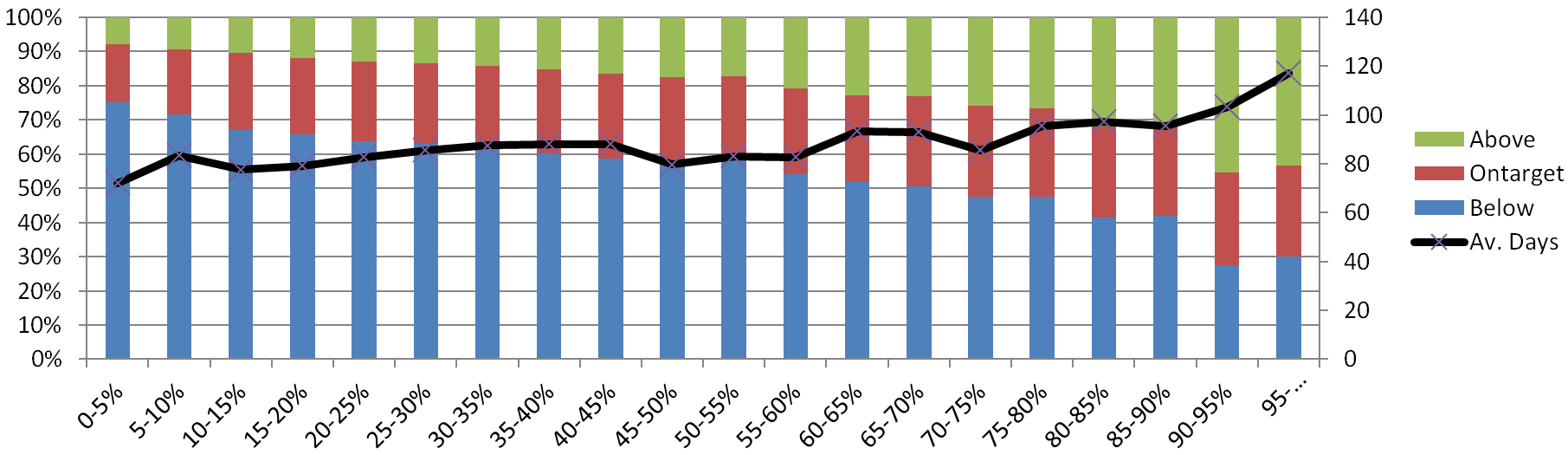}
\caption{4,251 users are classified using the token-based SVM (see text) and then sorted based on the distance from the decision boundary. 0-5\% refers to the 5\% of users least likely to be labeled ``above'', similarly 95-100\% refers to the 5\% of users \emph{most} likely to be labeled ``above''. The stacked plot shows the macro-averaged distribution across the logged days. The black line shows the average number of days logged for user in a given percentile group.}
\label{fig:mfp_users}
\end{figure}

The classification performance was sufficient though the category-based model did not perform better than the token-based model. We also performed an analysis to look at when the classifier errs. For both feature sets, both the false positive and false negative instances furthest from the decision boundary had less than 50\% in their modal class (``below'' or ``above''). In other words, their ground truth label had a low degree of confidence. Similarly, in all cases the individual instances furthest from the decision boundary had the correct labels assigned. This can also be seen in Figure~\ref{fig:mfp_users} where the fraction of ``above'' days for users increases from left to right, i.e., from being classified most strongly as ``below'' to being classified most strongly as ``above''. Interestingly, the graph also shows that users with a higher fraction of ``above'' days also tend to have logged more days in the system.

\subsection{Feature Analysis}\label{sec:feature_analysis}
Our main motivation was \emph{not} to predict if a user will be mostly above or below their weight goal, but rather, to understand the potential effect that different food choices might have on this outcome. To achieve this, we performed a feature analysis for the learned classification models. As we used a linear kernel, each feature is assigned a weight which can be directly interpreted with large and positive feature weights being indicate of ``above'' and large (in absolute value) and negative feature weights being indicative of ``below''.

\begin{table}[ht]
\centering
\caption{The 10 most discriminative features in the token-based linear SVM model. For each token, the foods logged by most users are listed. \label{tab:indicator_tokens}}{
\begin{tabular}{|c|c||c|c|}
\hline
   \multicolumn{2}{|c|}{Over Weight Goal} & \multicolumn{2}{|c|}{Under Weight Goal} \\ \hline
   Token             & Example Dish &   Token & Example Dish \\ \hline
oil & oil - olive                & cup & strawberries 1 cup${}^{*}$ \\
 wine & wine - table, red         & kroger & sugar - kroger${}^{*}$ \\
 added & quick added calories     & banana & turbana - banana \\
butter & salted butter${}^{*}$   & grapes & grapes - raw \\
 dairy & butter - dairy${}^{*}$   & almond & almond milk${}^{*}$\\
 original & original ranch${}^{*}$& value & value fries${}^{*}$\\
 pieces  & walnut pieces${}^{*}$  & egg & whole egg${}^{*}$ \\
 container&mayonnaise container${}^{*}$ & dole & dole banana${}^{*}$ \\
 lemon   & lemon juice${}^{*}$    & weight & weight control oatmeal${}^{*}$ \\
 mcdonalds & mcdo hash brown${}^{*}$ & breast & turkey breast meat \\ \hline
\end{tabular}}
\end{table}

Table~\ref{tab:indicator_tokens} shows the features with the 10 most positive (negative) weights on the left (right). Tokens such as ``oil'', ``butter'', or ``mcdonalds'' are all indicative for consuming more calories than planned. The ``added'' token in third position is mostly derived from using the ``quick added calories'' functionality. This functionality allows users to manually enter a summed caloric amount without having to enter each food item separately. Fruit tokens such as ``banana'', ``grapes'', or ``lemon'' are all indicative of staying below one's calorie goal.

\begin{table}[ht]
\centering
\caption{The 10 most discriminative categories in the category-based linear SVM model. For each category, the foods logged by most users are listed. A ``..'' indicates an omitted level for very long and multi-level categories\label{tab:indicator_categories}}{
\begin{tabular}{|c|c||c|c|}
\hline
    \multicolumn{2}{|c|}{Over Weight Goal} & \multicolumn{2}{|c|}{Under Weight Goal} \\ \hline
    Category             & Example Dish       &   Category & Example Dish \\ \hline
  beverage:alcohol & sabras - hummus      & meat:..:turkey & sliced turkey${}^{*}$ \\
  dessert:cake & cheesecake               & fruit & bananas - raw \\
  preparation:fry & eggs - fried${}^{*}$ & meat & turkey breast meat \\
  staple:wheat:pizza & pepperoni pizza${}^{*}$ & egg\textunderscore{dairy} & eggs - fried${}^{*}$ \\
  meat:pork & ham - sliced${}^{*}$ & meat:poultry & chicken breast${}^{*}$ \\
  dessert & cookies${}^{*}$ & dessert:..:caramel & caramel - caramels \\
  staple:other\textunderscore{cereal} & fiber one bar${}^{*}$  & fruit:..:banana & bananas - raw \\ 
 ..:..:coconut\textunderscore{oi}l & coconut oil${}^{*}$ & ..:milk\textunderscore{substitutes} & almond milk${}^{*}$ \\
  staple:root\textunderscore{and}\textunderscore{tuber} & potatoes - baked${}^{*}$ & preparation:bake & potatoes - baked${}^{*}$ \\
  staple:wheat:bread  & bread - italian  & snack:snack:donut & glazed donut${}^{*}$ \\ \hline
\end{tabular}}
\end{table}

Table~\ref{tab:indicator_categories} summarizes the category-based classification model. Overall, categories related to fruit, poultry, and baked foods are indicative of staying below one's calorie goals, whereas wheat, pork, and fried foods point towards going over.

It is interesting to see that desserts in general (denoted by the main category ``dessert'') are associated with logging too many calories, but caramels (denoted by the specific entity ``dessert:confectionery:caramel'') are associated with logging \emph{less} categories than one's goal. However, the average usage of caramels corresponds to only 130kcal, compared to 173kcal for any logging entry under ``dessert'' and 195kcal for generic cakes (``dessert:cake''). The appearance of donuts (``snack:snack:donut''), with an average of 180 kcal, in the under-goal class is also unexpected. This may be potentially caused by the collinearity of certain features although the regularization term in the SVM usually deals with this.

Note that Table~\ref{tab:indicator_categories} shows that ``sabras - hummus'' is incorrectly categorized as beverage:alcohol in our taxonomy (see Section~\ref{sec:data:diaries_taxonomy}). A false positive is caused by the token ``sabras'' which was incorrectly matched to ``Sabra'', a liqueur produced in Israel, contained in the beverage category. 
Example foods in Table~\ref{tab:indicator_tokens} \& \ref{tab:indicator_categories} that are marked with ${}^{*}$ were shortened to fit the table. The full list of names can be found in a footnote.\footnote{The full names were ``butter - salted butter'', ``butter - 1 pat - dairy'', ``hidden valley - original ranch'', ``walnuts - walnut pieces'', ``hellmann's - real mayonnaise 30fl oz container'', ``lemon juice - raw'', ``mcdonald's - hash brown from mcdonalds'', ``strawberries - raw, 1 cup'', ``light brown sugar - kroger'', ``almond milk - almond milk - vanilla - unsweetened'', ``wendy's - value french fries'', ``eggs - fried (whole egg)'', ``dole banana - bananas'', ``quaker oats - weight control instant oatmeal maple \& brown sugar'', ``eggs - fried (whole egg)'', ``little caesars - pepperoni pizza'', ``ham - sliced, extra lean'', ``cookies - chocolate chip, soft-type'', ``fiber one - fiber one bar, oats \& chocolate'', ``spectrum - coconut oil, unrefined'', ``potatoes - russet, flesh and skin, baked'', ``turkey, deli sliced - turkey'', ``eggs - fried (whole egg)'', ``protein - tyson chicken breast'', ``drinks - almond bilk (vanilla)'', ``potatoes - russet, flesh and skin, baked'', ``original glazed donut - krispy kreme''.}

% DISCUSSION
\section{Discussion}
%\note[Ingmar]{@AEK: Feel free to expand/edit this section}
The analysis shown in this paper is preliminary in parts but still serves to show the value of using food diaries for studying dieting success in real-world settings. More complex machine-learning models could be used to, e.g., look at the \emph{interaction} between food types. Conveniently, MFP provides this nutritional breakdown for the logged meals (see Figure~\ref{fig:mfp_screenshot}). Such analysis could shed light on the success of dieting practices that advocate the separation of carbohydrates and protein or similar approaches.\footnote{There are various types of ``food combining'' diets with the Hay Diet being one of the most prominent ones, despite the lack of success shown in randomized trials \cite{golayetal00ijo}.}

Our definition of whether a user is below or above their calories goal (see Section~\ref{sec:data:labeling}) is admittedly simple. For example, a user who is over their goal by 100\% for one day, but then under for 10\% for ten days would be labeled as ``under'' instead of ``on-target''. In fact, there are plausible alternative definitions but we do not expect them to change the results dramatically. For example, we had initially used a +/- 20\% margin in \emph{both} directions, not only towards below, and this gave similar list of discriminative tokens (Table~\ref{tab:indicator_tokens}).

More fundamentally, it is very difficult to determine if a shorter-than-usual log entry indicates a day of ``food abstinence'' or just an incomplete diary. Though we did not use it in this study, we could ideally obtain the weight loss goal, encoded in a picture posted in the profile page of some users. Having this information would also be helpful in distinguishing the small fraction of users on MFP who might be trying to \emph{gain} or \emph{maintain} weight rather than losing it. These users are not treated properly by our methodology though the analysis of ``which food is linked to being above the weight goal'' still holds.

Looking at the temporal patterns across a user's lifetime in the system, we did some preliminary analysis to see if users stopped logging food because of (i) achieving a set weight loss goal or (ii) getting frustrated by the failure to do so. For this, we looked at the fractions of users in a given ``temporal percentile range'', referring to a user's logging events buckets in 10\% of their total time range. For each temporal bucket, we then assign it the modal label during that period. In aggregate, over a user's lifetime in the system their daily success ratios changed only slightly, from an initial below-vs.-on-target-vs.-above of 61\%-23\%-16\%, to 63\%-20\%-17\% for the penultimate 80-90\% bucket. Only for the final 10\% of logging events, the distribution changed to 69\%-16\%-15\%. We are still unsure if this indicates (i) being most likely to ``over-perform'' by staying well below one's calories goal just before abandoning, or (ii) logging in a more and more incomplete manner at the end. 

Weight control is related to controlling both calorie intake and energy expenditure. In this analysis, we only looked at the former of these. However, given that platforms such as MFP also provide a way to log the latter, we deem this worthwhile for future exploration.

Generally, having more access to user profile information could help predict ``what type of diet will work for whom''. What works could depend both on what type of foods a user has access to (e.g., due to income, geography, or working hours) but also relate to general lifestyles (e.g., with increased peer pressure when eating in a group). Automatically generated, \emph{personalized} weight loss programs will definitely attract more attention in the future.

%quick added, dilligent, no short-cuts, ...

% CONCLUSION
\section{Conclusion}
This paper presents a study that uses public food diaries of more than 4,000 long-term active MFP users. Our analysis is centered around a classifier that, given a list of foods consumed by a person, predicts if they will be below or above their self-defined calorie goal. While certain findings are expected (``oil'' and ``mcdonalds'' being indicative of consuming too many calories) others are less obvious (poultry is linked to staying within one's goals, whereas pork indicates going above). 
Our results prove the feasibility of mining such data for health-related analysis. Especially with additional links to users' activity and general lifestyle patterns, automatically generated personalized and adaptive dieting seem a promising avenue to pursue. 
Health informatics is only starting to use the veritable gold mine that comes with public quantified-self data. This paper contributes to advances in this field by exploring how public food diaries can be mined to understand differences in unsuccessful and successful diets.

%Incorporate into system, e.g. for adaptive dieting ...

\section*{Acknowledgment}
% We would like to thank Living Analytics Research Centre (LARC), and Prof.\ Ee-Peng Lim in particular, for supporting a research visit by Dr.\ Weber in early 2015. This laid the foundation for this collaboration. 
%\note[Ingmar]{@AEK: Anybody else to acknowledge? Any grants?}

This work is supported by the National Research Foundation under its International Research Centre @ Singapore Funding Initiative and administered by the IDM Programme Office. In particular, we would like to thank Prof. Ee-Peng Lim for supporting a research visit by Dr.\ Weber in early 2015. This laid the foundation for this collaboration.

\bibliographystyle{abbrv}
\bibliography{mfp_food_diaries}

\begin{thebibliography}{10}

\bibitem{abbaretal15chi}
S.~Abbar, Y.~Mejova, and I.~Weber.
\newblock You tweet what you eat: Studying food consumption through twitter.
\newblock In {\em Conference on Human Factors in Computing Systems (CHI)},
  pages 3197--3206, 2015.

\bibitem{burkeetal11jada}
L.~E. Burke, J.~Wang, and M.~A. Sevick.
\newblock Self-monitoring in weight loss: A systematic review of the
  literature.
\newblock {\em Journal of the American Dietetic Association}, 111:92–--102,
  2011.

\bibitem{culotta14chi}
A.~Culotta.
\newblock Estimating county health statistics with twitter.
\newblock In {\em Conference on Human Factors in Computing Systems ({CHI})},
  pages 1335--1344, 2014.

\bibitem{golayetal00ijo}
A.~Golay, A.-F. Allaz, J.~Ybarra, P.~Bianchi, S.~Saraiva, N.~Mensi, R.~Gomis,
  and N.~de~Tonnac.
\newblock Similar weight loss with low-energy food combining or balanced diets.
\newblock {\em International Journal of Obesity}, 24(4):492--496, 2000.

\bibitem{halletal09sigkdd}
M.~A. Hall, E.~Frank, G.~Holmes, B.~Pfahringer, P.~Reutemann, and I.~H. Witten.
\newblock The {WEKA} data mining software: an update.
\newblock {\em {SIGKDD} Explorations}, 11(1):10--18, 2009.

\bibitem{svmlight}
T.~Joachims.
\newblock Making large-scale {SVM} learning practical.
\newblock In B.~Sch{\"o}lkopf, C.~Burges, and A.~Smola, editors, {\em Advances
  in Kernel Methods - Support Vector Learning}, chapter~11, pages 169--184. MIT
  Press, Cambridge, MA, 1999.

\bibitem{kraschnewskietal10ijo}
J.~L. Kraschnewski, J.~Boan, J.~Esposito, N.~E. Sherwood, E.~B. Lehman, D.~K.
  Kephart, and C.~N. Sciamanna.
\newblock Long-term weight loss maintenance in the united states.
\newblock {\em International Journal of Obesity}, 34:1644--1654, 2010.

\bibitem{kuebleretal13pone}
M.~Kuebler, E.~Yom-Tov, D.~Pelleg, R.~M. Puhl, and P.~Muennig.
\newblock When overweight is the normal weight: An examination of obesity using
  a social media internet database.
\newblock {\em {PLOS} {ONE}}, 8:e73479, 2013.

\bibitem{lietal14group}
V.~Li, D.~W. McDonald, E.~V. Eikey, J.~Sweeney, J.~Escajeda, G.~Dubey,
  K.~Riley, E.~S. Poole, and E.~B. Hekler.
\newblock Losing it online: Characterizing participation in an online weight
  loss community.
\newblock In {\em Conference on Supporting Group Work ({GROUP})}, pages 35--45,
  2014.

\bibitem{lindeetal15cct}
J.~A. Linde, R.~W. Jeffery, S.~J. Crow, K.~L. Brelje, C.~R. Pacanowski, K.~L.
  Gavin, and D.~J. Smolenski.
\newblock The tracking study: description of a randomized controlled trial of
  variations on weight tracking frequency in a behavioral weight loss program.
\newblock {\em Contemporary Clinical Trials}, 40:199--211, 2015.

\bibitem{meyeretal14chi}
J.~Meyer, S.~Simske, K.~A. Siek, C.~G. Gurrin, and H.~Hermens.
\newblock Beyond quantified self: Data for wellbeing.
\newblock In {\em Conference on Human Factors in Computing Systems (CHI)},
  pages 95--98, 2014.

\bibitem{parketal16cscw}
K.~Park, I.~Weber, M.~Cha, and C.~Lee.
\newblock Persistent sharing of fitness app status on twitter.
\newblock In {\em omputer-Supported Cooperative Work and Social Computing
  ({CSCW})}, page to appear, 2016.

\bibitem{payne15jmirmu}
H.~E. Payne, C.~Lister, J.~H. West, , and J.~M. Bernhardt.
\newblock Behavioral functionality of mobile apps in health interventions: A
  systematic review of the literature.
\newblock {\em JMIR Mhealth Uhealth}, 3:e20, 2015.

\bibitem{pellegmoore00icml}
D.~Pelleg and A.~W. Moore.
\newblock X-means: Extending k-means with efficient estimation of the number of
  clusters.
\newblock In {\em Conference on Machine Learning ({ICML})}, pages 727--734,
  2000.

\bibitem{rooksbyetal14chi}
J.~Rooksby, M.~Rost, A.~Morrison, and M.~C. Chalmers.
\newblock Personal tracking as lived informatics.
\newblock In {\em Conference on Human Factors in Computing Systems (CHI)},
  pages 1163--1172, 2014.

\bibitem{rusinetal13ijmi}
M.~Rusin, E.~Arsand, and G.~Hartvigsen.
\newblock Functionalities and input methods for recording food intake: A
  systematic review.
\newblock {\em International Journal of Medical Informatics}, 82:653--664,
  2013.

\bibitem{sciamanna11ajpm}
C.~N. Sciamanna, M.~Kiernan, B.~J. Rolls, J.~Boan, H.~Stuckey, D.~Kephart,
  C.~K. Miller, G.~Jensen, T.~J. Hartmann, E.~Loken, K.~O. Hwang, R.~J.
  Williams, M.~A. Clark, J.~R. Schubart, A.~M. Nezu, E.~Lehman, and
  C.~Dellasega.
\newblock Practices associated with weight loss versus weight-loss maintenance.
\newblock {\em American Journal of Preventive Medicine}, 41:159--166, 2011.

\bibitem{tengetal12websci}
C.-Y. Teng, Y.-R. Lin, and L.~A. Adamic.
\newblock Recipe recommendation using ingredient networks.
\newblock In {\em Web Science Conference ({WebSci})}, pages 298--307, 2012.

\bibitem{wagneretal14epj}
C.~Wagner, P.~Singer, and M.~Strohmaier.
\newblock The nature and evolution of online food preferences.
\newblock {\em EPJ Data Science}, 3(1), 2014.

\bibitem{wagneretal14www}
C.~Wagner, P.~Singer, and M.~Strohmaier.
\newblock Spatial and temporal patterns of online food preferences.
\newblock In {\em World Wide Web Conference (WWW)}, pages 553--554, 2014.

\bibitem{westetal13www}
R.~West, R.~W. White, and E.~Horvitz.
\newblock From cookies to cooks: insights on dietary patterns via analysis of
  web usage logs.
\newblock In {\em World Wide Web Conference (WWW)}, pages 1399--1410, 2013.

\end{thebibliography}

\end{document}